\documentclass[]{aa} 
\usepackage{graphicx}
\usepackage{txfonts}
\usepackage{longtable,lscape}
\begin{document}
   \title{Distinguishing post-AGB impostors in a sample of  pre-main sequence stars}

   \author{
             Rodrigo G. Vieira \inst{1}
           \and
              Jane Gregorio-Hetem \inst{1}
           \and
              Annibal Hetem Jr.\inst{2}
           \and
              Gra\.zyna Stasi\'nska \inst{3}
           \and
              Ryszard Szczerba \inst{4}
          }

   \institute{
              Universidade de S\~ao Paulo, IAG - Rua do Mat\~ao, 1226, 05508-900 - S\~{a}o Paulo, SP, Brazil 
              \email{vieira@astro.iag.usp.br}
         \and
              Universidade Federal do ABC, CECS, Rua Santa Ad\'elia, 166, 09210-170 Santo Andr\'e, SP, Brazil
         \and
              Observatoire de Meudon, LUTH, 5 Place Jules Janssen, 92190, Meudon, France
         \and 
              N. Copernicus Astronomical Center, Rabia\'nska 8, 87-100, Toru\'n, Poland        
          }


 
  \abstract
{A sample of 27 sources, catalogued as pre-main sequence stars by the Pico dos Dias Survey (PDS),  
is analyzed to  investigate a possible contamination by  post-AGB stars. The far-infrared excess, due to dust present in the 
circumstellar envelope, is typical for both  categories:   young stars and objects that have already left the main sequence and are 
suffering a severe mass-loss.}
{The presence of two known post-AGB stars in our 
sample inspired us to seek for other {{\it very likely}} or {{\it possible}} post-AGB objects among PDS  sources previously 
suggested to be Herbig Ae/Be stars,  by revisiting the observational database of this  sample.}
{In a comparative study with well known post-AGBs, several characteristics were evaluated: (i) parameters related to the circumstellar 
emission; (ii) spatial distribution to verify the background contribution from dark clouds; (iii) spectral features, and
(iv) optical and infrared colors. }
 {These characteristics suggest that 7 objects of the studied sample are very likely post-AGBs, 5 are possible post-AGBs, 
 8 are unlikely post-AGBs, and the nature of 7 objects remains unclear.}
{}


\authorrunning{Vieira et al.}
\titlerunning{Post-AGB impostors in a sample of pre-main sequence stars}

\maketitle
%

\section{Introduction}

   In the study of large samples of stars having unconfirmed nature,  the criteria  used in selection of candidates 
often cause sample  contamination with objects of other nature than that of  interest.
This is the case of recurring confusion between  pre-main sequence (pre-MS)
 stars,  and   post-asymptotic giant branch (post-AGB) stars. In spite of their totally different   
 evolution, both categories of objects share common characteristics. The observed  infrared (IR) excess, which  originates from
 circumstellar dust, can be explained by re-emission of thermal radiation produced by the central source in both cases.

Instigated by the presence of two confirmed post-AGBs in a sample of possible Herbig Ae/Be (pre-MS stars of intermediate mass),
we decided to analyze in detail  the objects having  spectral energy distribution (SED)  similar to those found in evolved stars. 
The sample was selected from the Pico dos Dias Survey (PDS)\footnote{PDS was a search for young stars based on infrared excess.} 
by choosing only the PDS sources showing SED more luminous in the near-IR than in the optical, 
which is also a known characteristic of post-AGBs.

Our goal is to distinguish, among the selected PDS sources, the {\it very likely}, the {\it possible} and the {\it unlikely} post-AGB objects, 
which could be included in ``The Toru\'n catalogue of Galactic post-AGB and related 
objects"\footnote{http://www.ncac.torun.pl/postagb2} compiled by Szczerba et al. (2007). 

We first briefly describe the circumstellar characteristics of both, young stars and post-AGB objects in order to review similarities 
and differences that are relevant for the present work.

\subsection{Pre-main sequence stars}

Direct imaging is the most reliable way to study the geometry of the envelopes, and to establish input disk parameters 
for SED fitting  models. 
However, sensitive imagery is constrained by several observational limits, being difficult to achieve for large samples.

In general, the circumstellar structure of  pre-MS stars is traced through indirect means like spectroscopic data, for example,
since the profile of spectral lines may be used to infer the physical conditions of line formation. 

Short-term spectral and polarimetric variability in Herbig Ae/Be (HAeBe) stars  indicate, for instance, circumstellar non-homogeneities
(Beskrovnaya et al. 1995), rotation, winds (Catala et al. 1999)  or magnetic field (Alecian et al. 2008). 
The observed SEDs of HAeBes can also be 
used to explain their IR excesses,  which are assumed to have originated from  a disk and/or an envelope 
surrounding the central star.
Different SED shapes have been used to classify the HAeBes according to the amount of IR-excess, which is one 
of the diagnostics of their evolutive phase in the pre-MS.

The classification schemes of HAeBes (Hillenbrand et al. 1992, Meeus et al. 2001) are based on the SED slope in the IR band,
related to the amount of IR excess and the geometric distribution of dust. However, these schemes do not consider more
embedded objects, which correspond to the first phases of the young stellar objects evolution.
A study of several other characteristics is required to check  the pre-MS nature of the
 candidates, since these embedded objects have SED quite similar to those classified as post-AGB.

\subsection{Post-AGB objects}

Following Szczerba et al. (2007), in order to refer to the evolutionary stage after the AGB phase, we prefer to adopt the general 
term post-AGB instead of proto-planetary nebula, since  {\it a priori}  we have no information if our objects could 
become Planetary Nebulae.
 Post-AGB stars are luminous objects with initial  mass between 0.8 and 8M$_{\odot}$. They completed their evolution in the asymptotic giant branch
 with a severe loss of mass (10$^{-7}$ to 10$^{-4}$ M$_{\odot}$/yr) (Winckel 2003).

When the mass-loss process  has removed  all the envelope around the central core, the AGB star suffers a transition from highly embedded
 object to a new  configuration of  detached envelope, being accompanied by changes in the SED shape, which acquires a double peak 
profile (Steffen et al. 1998). 
Van der Veen et al. (1989) classified a sample of post-AGBs, suggesting four classes according to the SED.
Classes I to III have
increasing SED slope from optical to far-IR wavelengths, while Class IV has double peak SED (maxima around near-IR and mid-IR).

Ueta et al. (2000, 2007) studied the characteristics of detached dust shells and different SED of post-AGBs by means of J-K and  K-[25] colors,
which  respectively describe the shape of the stellar spectrum in the near-IR, and the relation between the stellar (near-IR) and dust (mid-IR) peaks. 
According to these authors, post-AGBs can be separated in two kinds of  morphologies: elliptical (SOLE) that seem to have  
an optically thin shell, with starlight passing through in all directions, and bipolar (DUPLEX) that have an optically thick circumstellar torus, 
where the starlight passes through only along the poles.  Results from the HST survey of post-AGBs presented
by Si\'odmiak et al. (2008)  support this dichotomy in the morphology of the nebulosities which was confirmed by Szczerba et al. (in preparation) for a large
sample of post-AGBs. These authors found again clear differences in 
near-IR colors of SOLE and DUPLEX objects, which are also correlated with  the SED shape:
post-AGB class IV objects (double peak) are SOLE type, while class II or III (single peak) are DUPLEX.

Similar differences in SED shape are also found in pre-MS stars, indicating if the starlight is scattered in all directions (double peak SED) 
or not (single peak) 
(Hillenbrand et al. 1992, Malfait et al. 1998, Meeus et al. 2001).

The main goal of the present work is to analyze a selected  sample of possible
HAeBes that is contaminated by the presence of post-AGBs. As described in Sect. 2.,
the sample was extracted from  the PDS catalogue  by selecting the sources that show SED shape similar to post-AGBs. 
Seven characteristics, typical of evolved objects, have been used to distinguish
them from the young stars in our sample. The analysis of these characteristics is presented as follows. In Sect. 3 we discuss 
the association with clouds in order to check (i) the effects of the interstellar medium in far-IR observed fluxes, and 
(ii) the occurrence of isolated objects. Spectral features and their relation with  circumstellar
 characteristics are discussed in Sect. 4. 
Optical and IR colors are evaluated and compared to known post-AGBs in Sect. 5.
The main results are summarized in Sect. 6, according to the criteria used to identify the most probable evolved objects, while concluding
remarks and perspectives of future work are presented in Sect. 7.

\begin{table*}[ht] 
\caption{List of studied stars}
\smallskip
\begin{center}
{\scriptsize
\begin{tabular}{lcccccccccccc}
\hline
\noalign{\smallskip}
PDS&	2MASS& ST &	E(B-V) & J-H & H-K & K-[25]& J-[18] & [9]-[18] &[12]-[25]&[25]-[60] & W$_{{\rm H}\alpha}$ & P \\
	   &			   &	   &	mag	  & mag &	mag&	mag&	mag &	mag	  &	mag	 &	mag	    &	\AA	                      &	\%   	\\			
\noalign{\smallskip}
\hline
\noalign{\smallskip}
018	&	05534254-1024006	&	B7	&	1.42	&	2.17	&	1.68	&	7.41	&	10.26	&	2.04	&	0.42	&	0.10	&	40	&	--	\\
027	&	07193593-1739180	&	B2	&	1.41	&	1.47	&	1.32	&	8.15	&	9.78	&	2.39	&	0.84	&	0.50	&	88	&	--	\\
037	&	10100032-5702073	&	B2	&	1.60	&	1.86	&	1.40	&	8.28	&	10.34	&	2.55	&	1.04	&	1.00	&	105	&	2.76	\\
067	&	13524285-6332492	&	B	&	1.56	&	1.59	&	1.49	&	6.23	&	8.56	&	1.51	&	0	&	-0.24	&	90	&	2.73	\\
141	&	12531722-7707106	&	?	&	--	&	2.32	&	1.82	&	7.95	&	11.06	&	2.43	&	0.84	&	0.22	&	40	&		\\
168	&	04305028+2300088	&	F0	&	2.14	&	1.74	&	1.19	&	6.51	&	8.53	&	1.87	&	0.19	&	-0.45	&	8	&	5.84	\\
174	&	05065551-0321132	&	B3	&	1.01	&	0.76	&	0.65	&	10.38	&	--	&	--	&	1.17	&	1.95	&	65	&	--	\\
193	&	05380931-0649166	&	B9	&	1.34	&	0.97	&	0.99	&	7.20	&	8.18	&	1.86	&	0.16	&	0.52	&	12	&	2.25	\\
198	&	05385862-0716457	&	F0	&	1.13	&	1.03	&	1.01	&	6.95	&	7.60	&	2.79	&	1.70	&	1.07	&	8.5	&	--	\\
204	&	05501389+2352177	&	B1	&	1.10	&	1.24	&	1.22	&	9.68	&	10.73	&	3.31	&	2.00	&	0.96	&	245	&	--	\\
207	&	06071539+2957550	&	B?	&	1.20	&	1.18	&	1.21	&	8.20	&	--	&	--	&	0.62	&	1.77	&	5	&	--	\\
216	&	06235631+1430280	&	B2	&	1.19	&	1.32	&	1.28	&	8.14	&	9.59	&	1.65	&	0.33	&	2.03	&	200	&	--	\\
257	&	07414105-2000134	&	A	&	0.89	&	1.20	&	1.21	&	7.81	&	--	&	--	&	0.35	&	1.77	&	15	&	--	\\
290	&	09261107-5242269	&	A	&	0.66	&	0.35	&	0.28	&	10.9	&	--	&	--	&	0.26	&	2.33	&	-7	&	1.74	\\
353	&	12222318-6317167	&	B5	&	0.99	&	1.31	&	1.17	&	8.28	&	9.68	&	1.97	&	0.45	&	2.00	&	200	&	2.17	\\
371	&	13473141-3639495	&	O9?	&	1.40	&	1.74	&	1.31	&	7.17	&	9.18	&	2.36	&	0.58	&	-0.1	&	40	&		\\
394	&	15351712-6159041	&	F0	&	0.44	&	0.51	&	0.52	&	11.45	&	11.17	&	3.34	&	1.51	&	-0.7	&	-2	&	2.25	\\
406	&	16050392-3945034	&	A5	&	0.40	&	1.08	&	1.05	&	8.33	&	9.19	&	1.95	&	0.79	&	1.94	&	13	&	3.2	\\
431	&	16545918-4321496	&	A0	&	0.52	&	0.17	&	0.16	&	13.88	&	--	&	--	&	2	&	2.29	&	8	&	0.23	\\
465	&	17451419-1756469	&	B	&	1.23	&	1.36	&	1.48	&	8.37	&	9.73	&	--	&	1.51	&	0.88	&	110	&	7.88	\\
477	&	18003031-1647259	&	B1	&	1.37	&	1.43	&	1.19	&	7.42	&	9.08	&	2.22	&	0.41	&	-0.01	&	120	&	1.4	\\
518	&	18273952-0349520	&	OB 	&	2.30	&	1.74	&	1.35	&	6.85	&	--	&	--	&	0.37	&	1.53	&	600	&	3.8	\\
520	&	18300616+0042336	&	F3	&	1.26	&	1.23	&	1.09	&	6.88	&	8.13	&	2.21	&	0.70	&	0.47	&	33	&	3.67	\\
530	&	18413436+0808207	&	A5	&	0.47	&	1.59	&	1.66	&	7.93	&	9.92	&	2.12	&	0.52	&	-0.07	&	28	&	11.06	\\
543	&	18480066+0254170	&	B0	&	1.79	&	0.48	&	0.40	&	8.91	&	8.93	&	3.82	&	1.92	&	0.41	&	0.8	&	1.14	\\
551	&	18552297+0404353	&	B0	&	1.92	&	1.53	&	1.26	&	8.16	&	9.49	&	2.15	&	0.57	&	0.57	&	50	&	11	\\
581	&	19361890+2932500	&	B1	&	0.89	&	1.98	&	1.72	&	8.58	&	10.82	&	--	&	1.33	&	0.74	&	200	&	12.22	\\
\hline
\end{tabular}
}
\end{center}
Columns description: (1)  PDS name; (2) 2MASS identification; (3) spectral type;
(4) B-V excess;  (5, 6, 7, 8, 9,10, 11) infrared colors; (12) H$_\alpha$ equivalent width (negative values represent absorption lines); (13) intrinsic polarization.
\end{table*}

\section{Sample  selection based on IR-excess}
The Pico dos Dias Survey (PDS\footnote{PDS was conducted at the 
{\it Observat\'orio do Pico dos Dias}, which is operated by the Laborat\'orio Nacional de Astrof\'isica/MCT, Brazil}) 
(Gregorio-Hetem et al. 1992, Torres et al. 1995, Torres 1999)
was a search for young stars, based on the far-IR colors of T Tauri stars (TTs). 
Among  the detection of more than 70 new TTs, PDS also provided a list of 108 HAeBe 
candidates published by Vieira et al. (2003). 

Sartori et al. (2010) classified the HAeBe PDS candidates according to the SED slope (spectral index) and the contribution 
of circumstellar emission to the total flux. They verified that 84\% of the 
studied PDS sources can be confidently considered as HAeBe, while the nature of several candidates remains uncertain.

In this Section we describe the criteria adopted to select the objects studied in
the present work, by using spectral index and the fraction of circumstellar luminosity as tracers of  IR-excess.

\subsection{Spectral Index}

The PDS HAeBe candidates were
separated by Sartori et al. (2010) into three groups, according to the SED slope between optical and mid-IR, 
measured by the spectral index $\beta{_1}=0.75 \log (F_{12}/F_{V})-1$ (Torres 1999). We have selected 27 of the PDS sources of unclear nature,
mainly those with high spectral index  ($\beta_1 >0.7$), which correspond to the most prominent IR excesses. For objects showing single peak SED, the spectral index is related to the level of circumstellar extinction.
Even if the SED is double-peaked, high values of the  $\beta_1$ index also indicate high levels of circumstellar emission (at 12 $\mu$m), fairly larger than the optical flux (V band), independently of the stellar temperature.

Among the selected  sources there are two previously known post-AGBs:  
Hen 3-1475 (PDS465) and IRAS19343+2926 (PDS581), 
 according to evidence reported in the literature 
(Riera et al. 1995, Rodrigues et al. 2003, Bowers \& Knapp 1989),
which motivated us to study their differences and similarities when compared to 
other objects of the sample.

The list of the selected objects is presented in Table 1, giving their identification and the parameters used
in the present work as spectral type (when available),  B-V excess, IR colors (using {\it 2MASS, AKARI} and {\it IRAS} data), and
equivalent width of H$\alpha$ line. Intrinsic polarization
obtained by Rodrigues et al. (2009) is also given, when available.

\begin{figure*}[]
\includegraphics[angle=0,width=7.0in]{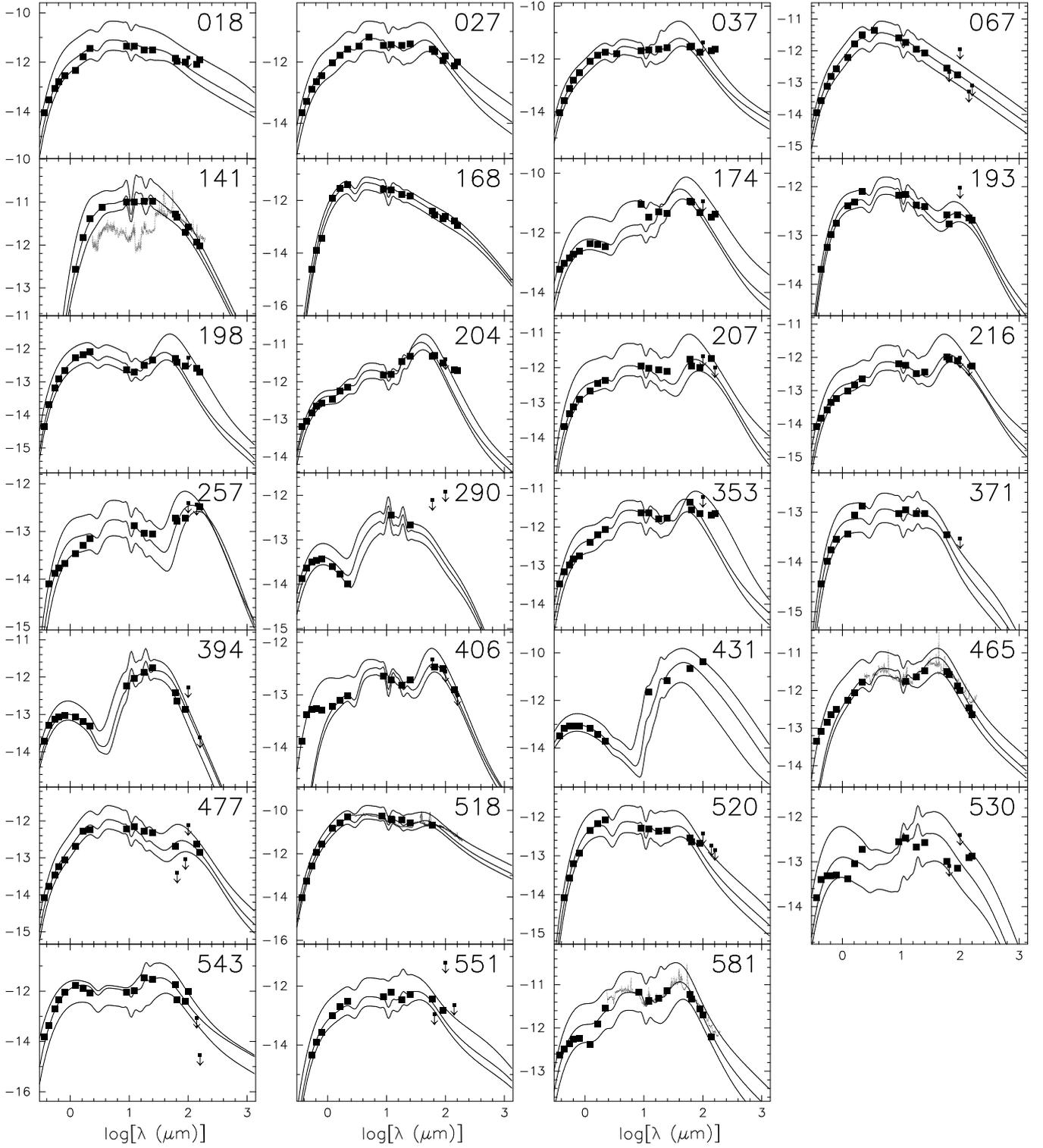}
\label{sed}
\caption{Observed SED of our sample showing log($\lambda$F$_{\lambda}$ in [Watt m$^{-2}$] {\it vs.} log($\lambda$) in [$\mu$m]. 
Filled squares represent optical photometry, {\it 2MASS}, {\it AKARI}, {\it MSX} 
and {\it IRAS} data, while dots are used to plot ISO spectra. Solid curves indicate the variation of  the calculated SED.}
\end{figure*}

\subsection{Fraction of Circumstellar Luminosity}

Different evolutive classes of pre-MS stars are defined according the observed SED, which is affected by several 
characteristics of circumstellar envelopes like: chemistry and size distribution of grains, and inclination of the system. 

The conspicuous grain features  around 3.1 $\mu$m and 10 $\mu$m can be checked through {\it ISO} or {\it Spitzer} data, in order to better infer the 
nature of the circumstellar matter. However, due to the lack of near-IR spectral data for our whole sample\footnote{ Only PDS141, 465, 
518 and 581 were observed by {\it ISO} or  {\it Spitzer}}, disk models cannot be constrained in the present work.
For this reason, we decided to adopt a simple model (Gregorio-Hetem \& Hetem 2002) for the sole purpose of estimating the fraction of circumstellar luminosity. 
Regardless of envelope or disk parameters, we are interested in determining the integrated observed flux, in order to estimate the 
contribution of circumstellar flux as a fraction of the total flux, defined by $f_{\rm Sc}$= (S$_{total}$-S$_{star}$)/S$_{total}$.  

Figure 1 shows the synthetic reproduction of observed SEDs of our sample, estimated from the blackbody emission of three
components: a central star (S$_{star}$), a flat passive disk (S$_{d}$), surrounded by a spherical envelope (S$_{e}$).
Different temperature laws are adopted: $ T_{\rm d} / T_{\rm star} \propto (r_{\rm d})^{-0.75}$ for the disk (Adams \& Shu 1986),
and $ T_{\rm e} / T_{\rm star} \propto (r_{\rm e})^{-0.4}$ for the envelope (Rowan-Robinson 1986),
as  used by Epchtein, Le Bertre \& L\'epine (1990) to reproduce the IR data of carbon stars.
The observational data are from {\it PDS} (optical photometry), {\it 2MASS, MSX, IRAS, AKARI} and {\it ISO} Catalogues. 
Three curves calculated by the adopted model, without fitting purpose, illustrate the expected variation 
on the fraction of circumstellar luminosity. In the case of PDS465, for example, $f_{\rm Sc}$
varies from 0.90 to 0.94, which is  the typical dispersion found in our sample, leading us to adopt
an error of  0.02 in the estimation of $f_{\rm Sc}$.

All objects of our sample have $\beta_1 > 0.7$ and  $f_{\rm Sc} > 0.7$ that correspond to 
large amounts of circumstellar emission, suggesting two possible scenarios: they are 
embedded pre-MS stars or they possible are  post-AGBs.

We are aware that it is important to check if the data points around 100 $\mu$m are reflecting the  
cloud and not the immediate vicinity of  the star, due to the large {\it IRAS} beam size in these bands, mainly for objects
lacking of {\it ISO} spectra or {\it AKARI} data, for example.
In  the next section we analyze  the background contribution from clouds, aiming to avoid a possible contamination 
of interstellar matter in the far-IR data.

\section{Association with dark clouds}
The star formation process is typically associated with  clouds, where the gravitational collapse of pre-stellar cores gives rise 
to the embedded protostars. Thus, the association  with these clouds may be an indication  of young-star nature. Unfortunately,
 only rough distance measurements are available for  our sample (Vieira et al. 2003), which makes uncertain the distinction between a true association 
from a projection effect. 
Due to this restriction, we decided to evaluate the {\it probability} of association with clouds, 
where the objects {\it apparently} closer to  dark clouds  {\it  probably} have young nature (age $<$  2 Myr), while
stars located far from clouds are expected to be evolved objects.

However, this cannot be used as a deterministic characteristic, since several examples of isolated pre-MS stars are known,
as AB Aur, HD 163296, and HD100546, for example  (e.g. van den Ancker 1999). 
Our main goals in investigating the spatial relation to dark clouds are twofold: (i) to verify the occurrence of isolated objects in 
our sample; and (ii) to estimate a possible far-IR  background contamination.

\begin{figure*}[t!]
\includegraphics[angle=0,width=2.5in]{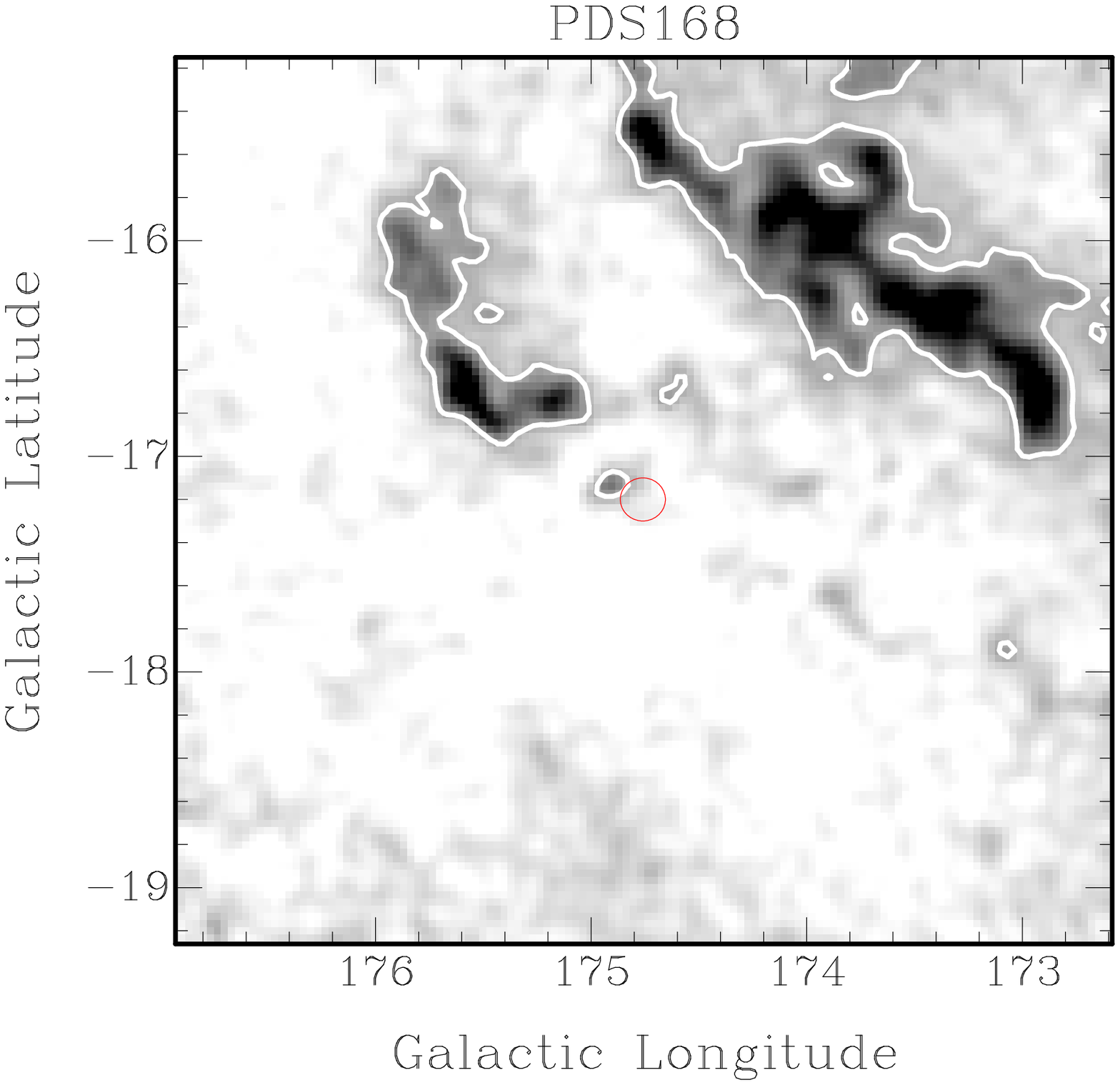}
\includegraphics[angle=0,width=2.5in]{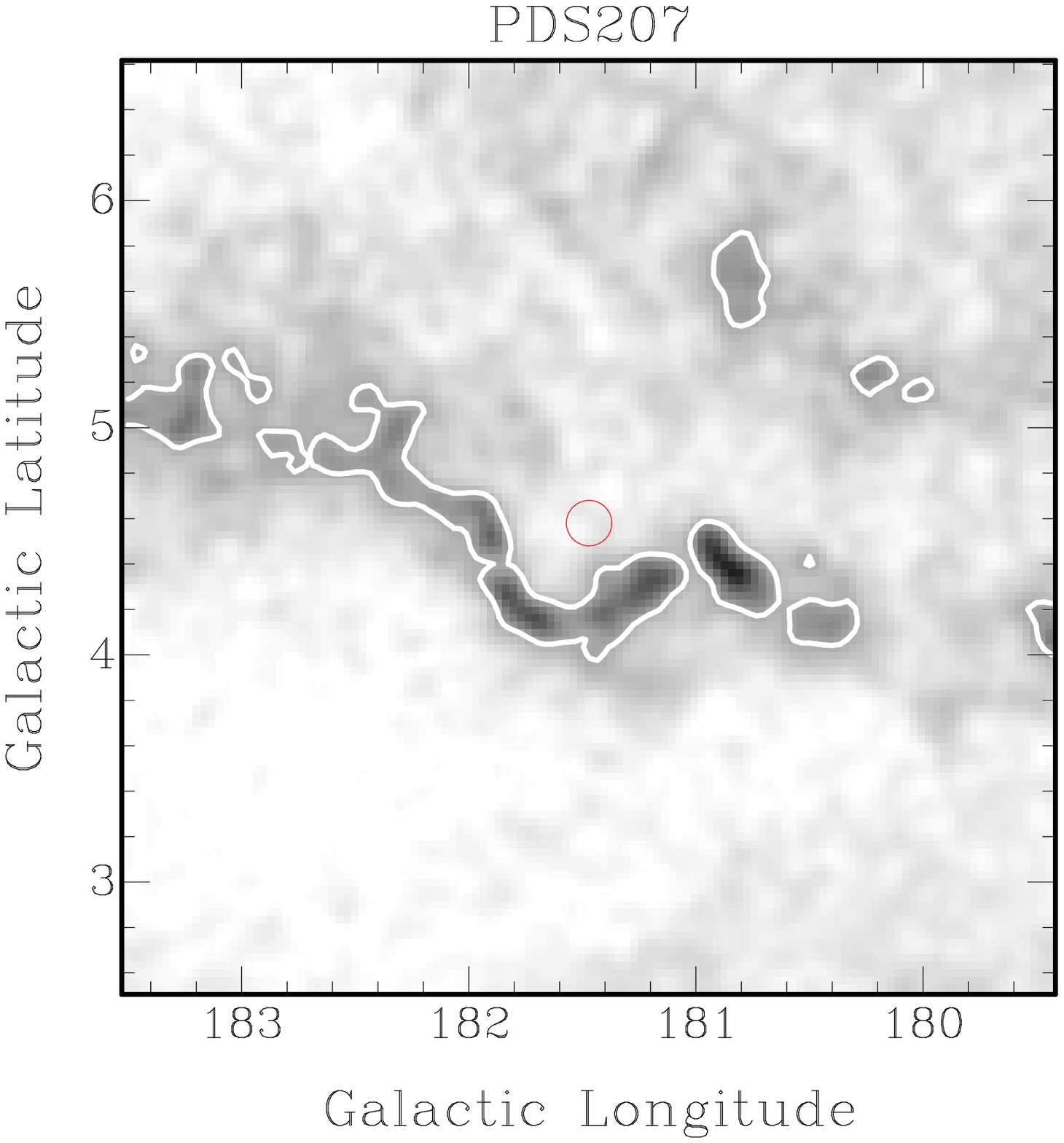}
\includegraphics[angle=0,width=2.5in]{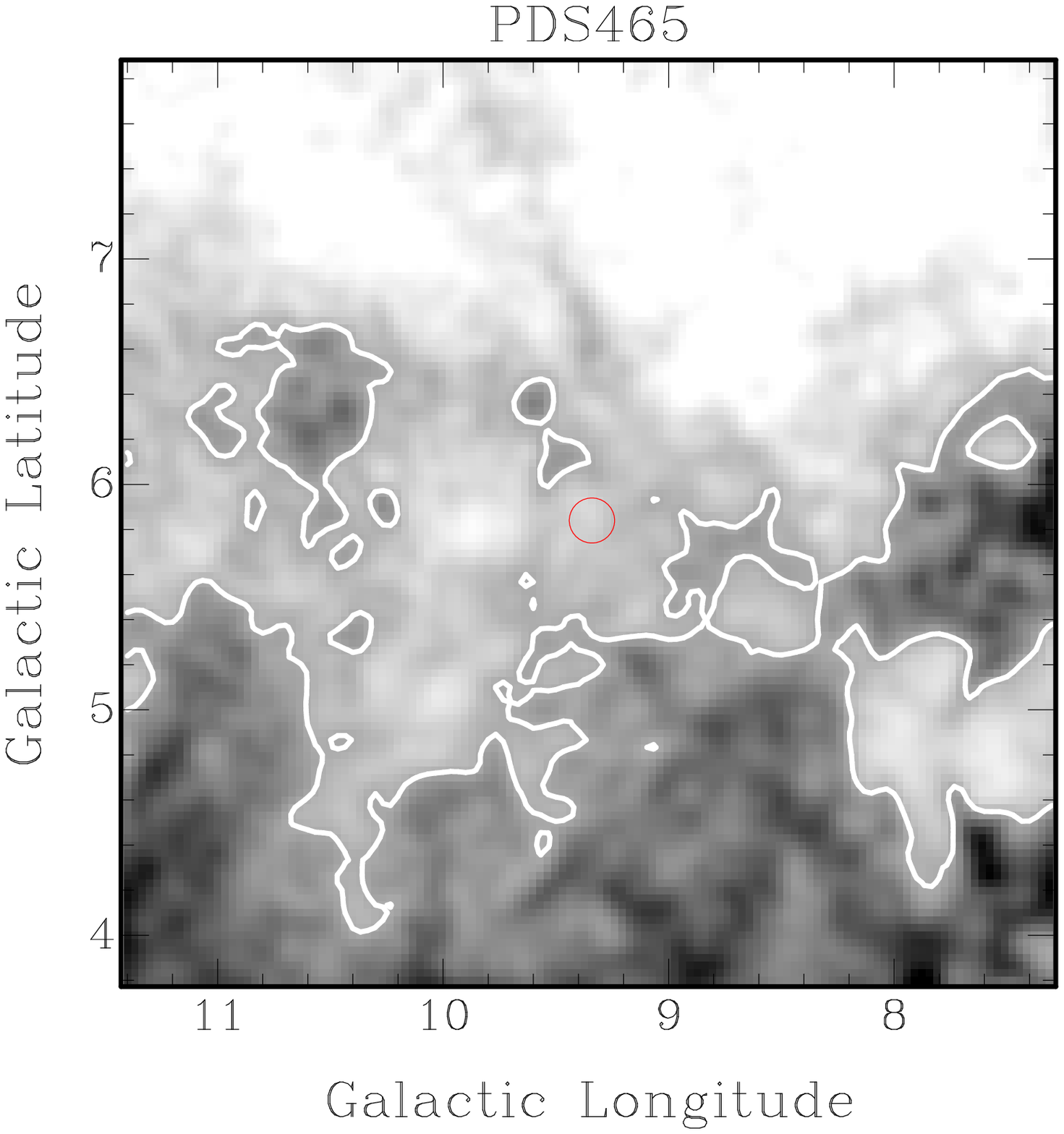}
\caption{Position of some of the objects (displayed by the central circle) in the Galactic extinction map produced by Dobashi
et al. (2005). The halftone grey color varies from 0 to 3 mag, while the contours show $A{_V}=1.0$ mag level.}
\label{Galactic extinction map}
\end{figure*}

\subsection{Distance to the edge of nearest clouds}

Aiming to infer the association with dark clouds, we have made use of the catalogue information compiled by Lynds (1962) and by
 Feitzinger \& St\"uwe (1984). 
These complementary works allowed us to find the dark clouds closer to the objects of our sample.
For each dark cloud of these catalogues is assigned an {\it opacity class} number, which ranges from 1 to 6. We have adopted the 
criterion of choosing  the closest dark cloud with opacity class 3 at least. This choice restricts the selected clouds
to those with relevant extinction level, although it must  be kept in mind that these opacity classes were defined based on visual
inspection of photographic plates. Given the selection of the closest clouds, we estimated their area and distance to each star of our
sample. These data have allowed us to roughly estimate the distance of our objects to the border of the selected clouds, defined by
 $d_{edge} = D - 0.5 ~\times ~a^{1/2}$, where D stands for the projected distance from the star to the cloud's center. By this way,
a negative value for $d_{edge}$ means that a star is found ``inside" the angular region enclosed by the cloud of  area {\it a}. 
A square area was adopted  in this first order  calculation that does not take into
account the actual shape of the cloud, which usually presents a filamentary geometry. Nevertheless, 
$d_{edge}$ quantifies a possible association with the  dark cloud.

\subsection{Reddening}

Since  the information obtained from catalogues of dark clouds is related to arbitrary opacities, we adopted  two other 
methods to evaluate dark cloud effects on the objects of our sample. First we used an extinction map, which quantifies
the cloud obscuration. In the second method we estimated the color excess in order to obtain the total extinction, due to  
both interstellar  (clouds) and circumstellar (disk/envelope) extinction. These two methods are described as follows.

\subsubsection{ Extinction Map}

We have adopted the Galactic map of visual extinction derived by Dobashi et al. (2005) using  the classical star count method on  the
 optical database of the Digitized Sky Survey I (DSS). The map\footnote{Available at http://darkclouds.u-gakugei.ac.jp}
 provides the value of the extinction $A_V$ for the region defined by $|b| \leq 40^{\circ}$.
 For each object, we have taken the $A_V$ value and its mean dispersion in a neighboring region of 1 square
 degree, aiming to quantify the fluctuation of this extinction level in the interstellar medium. This quantity reveals the presence of a
possible progenitor dark cloud behind the selected object. 

The obtained extinctions and respective dispersions are given
in Table 2. For illustration, some specific regions of Dobashi et al.'s map are shown in Figure 2.
It must be stressed that a possible association of our objects with clouds may be due to projection effects, where the dark cloud and 
the star are just on the same line of sight but not  physically associated. However, since we do not know the distances of our objects,
we can only speak in terms of {\it probability} of association. Objects having low values of $A_V$ probably are isolated, 
but other methods are required to confirm this information.

\subsubsection{Color Excess}

The visual extinction may be also estimated by means of the color excess that is caused by the absorption and scattering of the 
emitted light, occurring not only in the ISM but also (and sometimes predominantly) in the circumstellar environment of our sample.
Aiming to estimate this total extinction, we evaluated  $E(V-I)$ instead of $E(B-V)$, since the color excess on  $V-I$ is less affected 
by the ultraviolet excess (Strom et al. 1975). The intrinsic colors were selected from Bessel et al. (1998) for different sets of
effective temperature and surface gravity. 

Based on the spectral type given in the {\it PDS} Catalogue, an estimative of effective temperatures was obtained by adopting the empirical 
spectral calibration provided by de Jager \& Nieuwennhuijzen (1987). 
Considering that the luminosity classes are uncertain for most objects of our sample, the color excess was estimated
by assuming two possible classes: main sequence and supergiant  that are adopted to represent HAeBes and post-AGBs  respectively.
Then, for each pair of luminosity class and $T_{eff}$ we established a $log~g$ value extracted from Straizys \& Kuriliene (1981), 
enabling the most convenient choice of intrinsic colors. The $E(V-I)$ values were converted to $E(B-V)$ by the relation derived by 
Schultz \& Wiemer (1975): $E(B-V)= E(V-I)/(1.60 \pm 0.03)$. A normal reddening 
$R_V=3.14 \pm 0.10$, which is a mean value of the interstellar extinction for several directions of the Galaxy studied by Schultz 
\& Wiemer (1975), was adopted to estimate the total extinction that includes both ISM and circumstellar effects, given by  
$A_{Vtot} = 3.14 ~\times~E(B-V)$. 
 These are minimum values, since  anomalous 
extinction ($R_V>$4, for example) may be found in dense interstellar clouds (e.g. Savage \& Mathis 1979). 
We estimate that  $A_{Vtot}$ can deviate by about 50\%.

\begin{figure*}[!t]
\includegraphics[height=7.2in,angle=270]{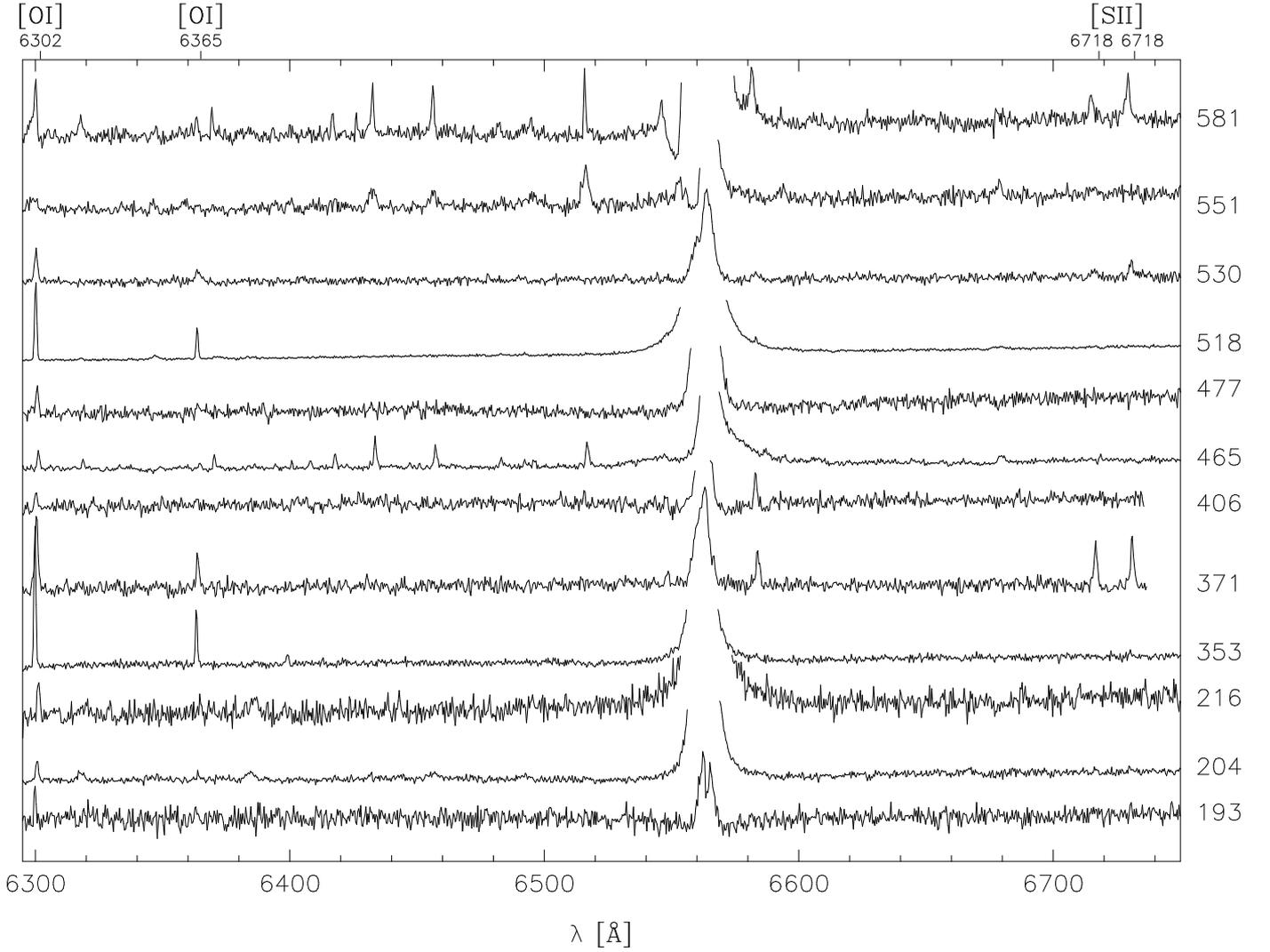}
\caption{Optical spectra of part of the sample, identified by the PDS number in the right side.  The H$\alpha$ feature is not shown in order to enhance details of
other emission lines. Intensity scale is arbitrary.} 
\label{spec}
\end{figure*}


\subsection{Background contribution}
Far-infrared emission may also reveal the contamination by interstellar material on the observed SED. Aiming to evaluate the effects
of background contamination, we have made use of {\it IRAS} images 
at 100 $\mu$m, which is dominated by the ISM cirrus emission.  For each {\it IRAS} image that contains one or more of the 
selected objects, we have chosen a region of 1 square degree in the neighborhood of the point sources, in order to estimate 
the background mean density flux and its standard deviation for each object. These background regions were selected avoiding the 
contamination of point sources in a radius of $1^{\circ}$. Table 2 gives the mean fluxes at 100$\mu$m ($<$F$>$) obtained in the 
background regions.

To quantify the excess of 100$\mu$m flux we compared the flux density of the point source (F$_{100}$)  with  the mean background 
flux explained above, defining the fraction $f_{\rm F100}$ = F$_{100}$/(F$_{100}$+$<$F$>$) as an indication of flux excess.
If $f_{\rm F100} \sim$ 0 the ISM dominates the local emission, while $f_{\rm F100} \sim$ 1 corresponds to an infrared excess mainly
due to the circumstellar contribution.

We also evaluated the fraction of ``net" visual extinction, which roughly corresponds to the optical depth of the circumstellar shell, 
by subtracting the interstellar obscuration ($A_V$) from the total extinction ($A_{Vtot}$).  In this case, we define 
$f_{\rm Av}$ =($A_{Vtot}$-$A_V$)/$A_{Vtot}$, were the interstellar extinction is obtained from the map of Dobashi et al.
(2005) and the total extinction is calculated from the color excess estimated from magnitudes and spectral types given in the PDS 
catalogue (see Sect. 3.2.2).
In this case, the effects of the adopted class of luminosity in the color excess estimation are negligible.

Table 2 shows the estimated values of $f_{\rm Av}$ and $f_{F100}$. Upper limits of $f_{F100}$ were estimated for sources
 having bad quality of 100 $\mu$m data. Excepting PDS518, all the sources with well determined $f_{F100}$ tend to show  
circumstellar extinction increasing with high levels of source emission.
The two known post-AGB objects of our sample have $f_{F100} >$ 0.5 and $f_{\rm Av} >$ 0.5, meaning that circumstellar material 
has respectively prominent infrared flux and optical depth, significantly above the levels found in the corresponding background.
We consider these high levels of  $f_{\rm Av}$ and $f_{F100}$ characteristics of post-AGBs.

\section{Spectral Features}

Optical spectra, obtained by the PDS 
team\footnote {The original files were made available by S. Vieira, on the website http://www.fisica.ufmg.br/$\sim$svieira/TRANSF/},
are shown in Figure 3 for the spectral regions  containing  [OI] (6302, 6365 {\AA})
and [SII] (6718, 6732 {\AA}) nebular lines, among others. 
The spectra of some sources are not displayed due to different reasons: they have no emission lines, excepting H$\alpha$ 
(PDS174, 198, 290, 394, 520, and 543);
they are too noisy  (PDS168, 207, and 431); the 6300-6500 {\AA} range was not covered (PDS18, 27, 37, 67, and 141), or the  
spectrum is not available (PDS257).  

Almost half of our sample shows the  [OI] 6302 {\AA} emission line, but only 
PDS371 shows features similar to those found in evolved objects, like planetary nebulae (Pereira \& Miranda 2005).
Considering the absence
or unclear detection of these features for the whole sample, they were not used to diagnose the nature of our objects.

The equivalent width of the H$_\alpha$ line (W$_{{\rm H}\alpha}$) of the objects in our sample was measured whenever possible and 
reported in  Table 2.  In the case of spectrum showing low signal-to-noise, we 
used the W$_{{\rm H}\alpha}$ published in the PDS catalogue.  

Most of the spectra show strong  H$_\alpha$ emission line, but two of them appear in absorption (PDS290 and 394). Line 
variability is a possible explanation to the weak H$_\alpha$, usually seen in pre-MS stars. 

A comparison of the fraction of circumstellar luminosity ($f_{\rm Sc}$) with the strength of H$_\alpha$ line is shown in Figure 4. The distribution of 
objects with  W$_{{\rm H}\alpha} <$ 50 {\AA} spreads along the whole range of $f_{\rm Sc}$. On the other hand, objects showing strong 
emission line are mainly concentrated in the region of $f_{\rm Sc}$ $>$ 0.87. As illustrated in Fig. 4, a similar distribution is found 
in the diagram of W$_{{\rm H}\alpha}$ in function of $f_{\rm Av}$, which is indicative of optical depth of the circumstellar shell, discussed 
in Sect. 3.3. 

\begin{figure}[!t]
\includegraphics[height=9cm,angle=270]{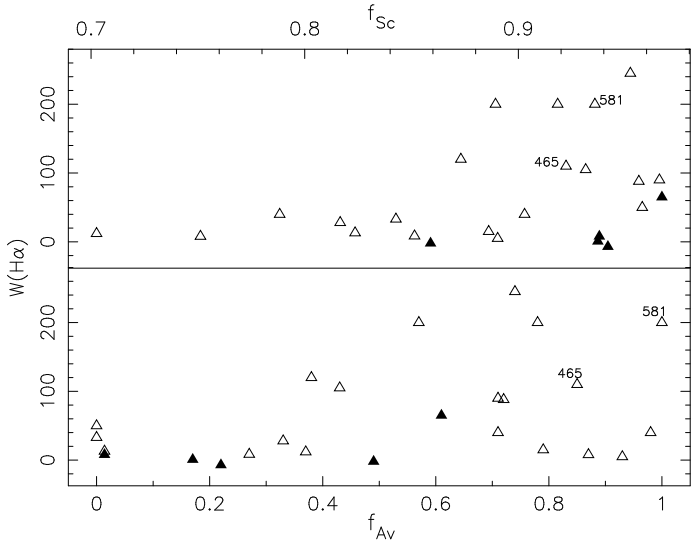}
\caption{Equivalent width of H$_{\alpha}$ line compared to $f_{\rm Sc}$ (top) and $f_{\rm Av}$ (bottom) for our
stars, excepting PDS518, which has W(H$_{\alpha}$)=600 \AA.  Filled triangles are used to show the PDS sources with double peak in
the SED, while open triangles represent single peak.} 
\label{ha}
\end{figure}

\section{Analysis of color-color diagrams}

To understand the nature of our objects  it is necessary to carry out a multi-wavelength analysis. For this purpose, 
we have made use of two-color diagrams in several spectral ranges. The photometric optical
data were obtained from the {\it PDS} catalogue. The measurements were taken in the Johnson-Cousins photometric system 
$UBV(RI)_c$, with the high-speed multicolor photometer (FOTRAP) described by Jablonski et al. (1994). 

The catalogues {\it 2MASS, IRAS, MSX}, and {\it AKARI} (when available)  provided the data for the infrared colors analysis of our objects.
Optical and near-IR photometric data were corrected for extinction based on the $A_{Vtot}$  calculated in Sect. 3.2.2,
adopting  the A$_{\lambda}$/A$_V$ relations from Cardelli et al. (1989).

\subsection{Optical colors}

The spectral region represented by the $B-V ~{\it vs.} ~U-B$ diagram is dominated  by the photospheric radiation, 
which makes it very interesting to study the nature of the central source. The optical colors diagram is shown in Figure 5 
displaying those of our objects that have available $UBV$ magnitudes.
For each object, two different pairs of colors are plotted according to the extinction correction explained 
in {\it Sect. 3.2.2} (by adopting two luminosity classes). Superimposed to this distribution are plotted the theoretical curves 
for $log~g$ = 2, 3, and 4, calculated by Bessel et al. (1998).

It is expected that HAeBes would have $log~g \simeq$ 4, similar to main sequence stars, whereas the post-AGB would have 
$log~g <$  2, typical of  supergiants.
However, as shown in Figure 5 the theoretical curves are clearly separated only for positive $B-V$ values, due to the presence
of the Balmer discontinuity, which is sensitive  to surface gravity in this spectral range, being more significant for supergiants. 
In this case, only PDS406 seems to satisfy the gravity criterion. 
On the other hand, for $B-V \lesssim 0$, where part of our objects are located, the  intrinsic colors from models 
for  $log~g$= 3 and 4 are similar. In this case,
we consider $U-B$ excess typical of evolved objects the values above the theoretical lines, as indicated for PDS465 and 581, the post-AGBs 
of our sample, as well as PDS174, 216, 353, 477, and 518.

\begin{figure}[!t]
\includegraphics[height=9cm,angle=270]{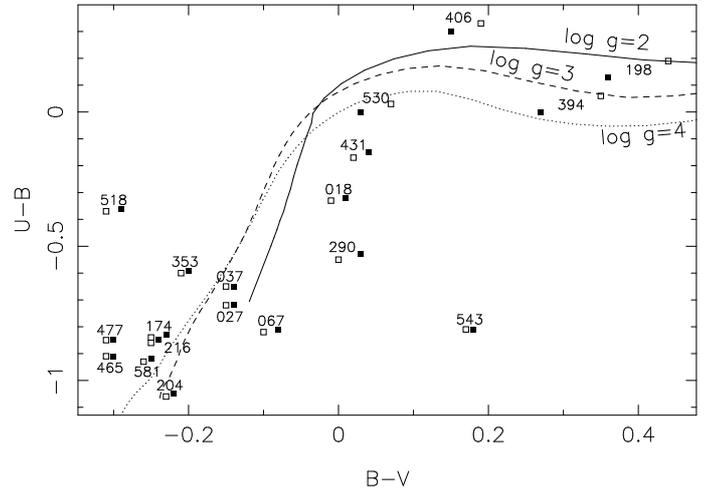}
\caption{Optical colors of the studied sample,  after dereddening using different luminosity classes: main sequence (filled squares) 
and supergiants (open squares). The intrinsic colors from Bessel et al. (1998) are displayed for different  surface gravities.}
\label{ubv}
\end{figure}

\subsection{Infrared colors}

The IR colors probe the thermal emission from grains that can be originated from dust at different temperatures: (i) cold grains 
found  in  circumstellar shells at  temperatures lower than  300K, or (ii) warm dust (T $\sim$ 1000K) produced by a recent episode 
of mass loss in the case of post-AGBs, or from a protostellar  disk in  pre-MS stars.

\subsubsection{Far-IR}

Diagrams of IR colors have been used to identify post-AGB candidates according to their expected {\it locus} that
is related to their evolution. Van der Veen \& Habing (1988) suggested an evolutive sequence in a diagram of IRAS colors for  post-AGB stars, 
defined by their mass-loss rate, in a scenario of short period Miras evolving into long period OH/IR stars. However, this hypothesis  is based on 
observational results that are not firmly supported since the distances to OH/IR stars are uncertain (Whitelock et al. 1991). Other authors 
(e.g. Feast \& Whitelock 1987) favor other scenario in which the Mira period is a function of its initial mass, where little changes in period or 
luminosity occur during the Mira evolution. This second scenario is supported by the results, obtained by Whitelock et al. (1991), on the 
evolution of Miras and the origin of their Period-Luminosity relationship.

Garc\'\i a-Lario et al. (1997), hereafter GL97, 
analyzed a sample of post-AGBs 
comparing their position in the [12]-[25] {\it vs.} [25]-[60] diagram with several categories of objects, as for example: OH/IR 
variables (Sivagnaman 1989); pre-MS stars (Harris et al. 1988); and compact  HII regions (Antonopoulos \& Pottasch 1987). 

\begin{figure}[!t]
\includegraphics[height=9cm,angle=270]{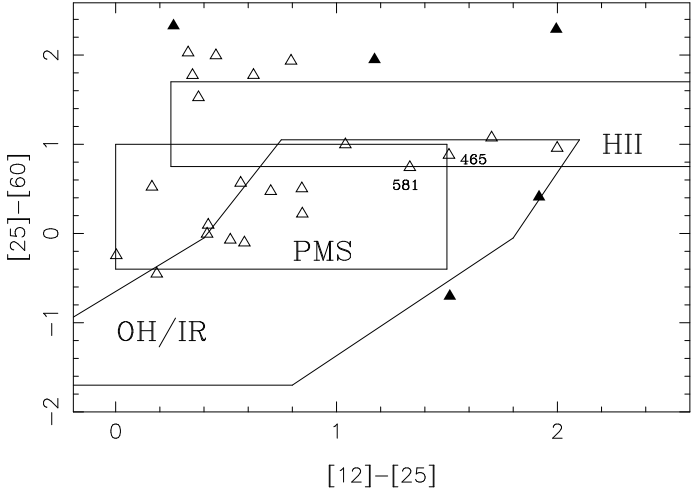}
\caption{Far-IR color-color diagram showing 
the regions for different categories of objects (GL97).
Filled triangles are used to show the PDS sources with double peak in
the SED, while open triangles represent single peak. 
} 
\label{iras}
\end{figure}

Figure 6 shows the distribution of our sample in the IRAS color-color diagram, with respect to the regions occupied by different classes of objects, 
as proposed by GL97. 
About 70\% of our objects are located in the left part ([12]-[25] $<$ 1) of this diagram, coinciding with the pre-MS box or above 
the HII region box  ([25]-[60] $\geq$ 1.5). 
Our remaining objects, among them PDS465 and 581, are found  approximately 
in the right side of the diagram ([12]-[25] $\geq$ 1). The 12 $\mu$m flux density distribution for IRAS sources located in this part of the diagram 
was studied by van der Veen et al. 
(1989, see region IV in their Figure 1). Based on the spatial distribution compared to the number of sources as a function of the 
12 $\mu$m flux density, they proposed a separation of objects in this region of the IRAS colours diagram. According to van der Veen
et al. (1989) young stars, which are associated to clouds, are those sources with F$_{12} < $ 2 Jy, while post-AGBs have F$_{12} > $ 2 Jy.

\begin{figure}[!t]
\includegraphics[height=9cm,angle=270]{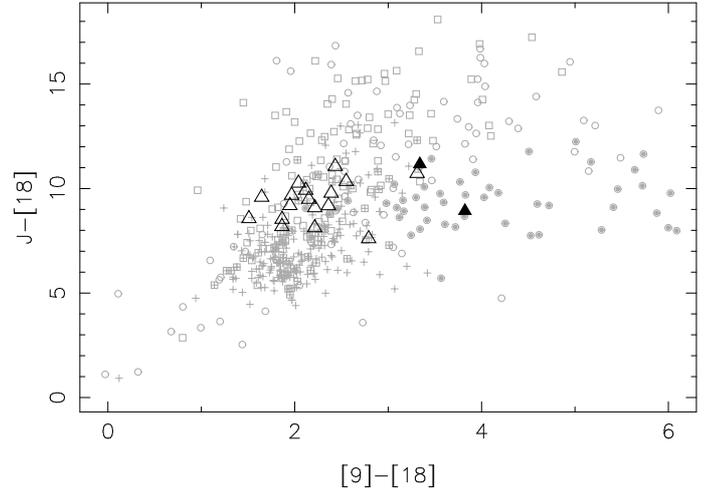}
\caption{Mid-IR color-color diagram based on {\it AKARI} fluxes. 
Triangles indicate our sample and grey circles represent the {\it very likely} post-AGBs from the Toru\'n Catalog.
Filled symbols are used for objects having double peak in the SED, while open symbols represent single peak.
The distribution of young objects is shown by grey squares (YSOs detected by {\it Spitzer}) and grey crosses 
(HAeBe stars from {\it HBC}). 
}
\label{akari}
\end{figure}

\subsubsection{Mid-IR}

The fluxes at 9, 18, 65, 90, 140, and 160 $\mu$m from the {\it AKARI} All Sky Survey (Ishihara et al. 2010) have been obtained 
for 78\% of our sample. These data were used in the SED fitting discussed in Sect. 2.2. Following Ita et al. (2010) we verified
the position of our sample in the $J-[18] ~{\it vs.} ~[9]-[18]$ diagram, expressed in Vega magnitudes. In Figure 7,
the {\it AKARI} colours  of our sources are compared 
with the distribution of the {\it very likely} post-AGB stars from the Toru\'n Catalogue (Szczerba et al. 2007). The position 
of young objects is also displayed in this diagram, by plotting the  YSOs from the {\it Spitzer} survey of young stellar clusters 
(Gutermuth et al. 2009) and the HAeBe stars from the  Herbig \& Bell Catalog ({\it HBC},~1988).
In spite of the better quality of the {\it AKARI} data, when compared to {\it IRAS} colours, the overlap of different categories also occurs 
in Figure 7, not distinguishing young objects nor evolved stars. Most of our sample is located in this overlap region, 
but three sources (PDS 204, 394, and 543)  coincide with the region of  post-AGBs, 
appearing more clearly separated, outside the main {\it locus} of the HAeBe stars, for example.

\subsubsection{Near-IR}

Near-IR photometry can be used to check whether the main source of emission is photospheric, nebular or due to  the presence 
of a circumstellar envelope. Figure 8 shows the  $J-H ~{\it vs.} ~H-K$ diagram studied by GL97 to compare AGB and
post-AGB stars with different 
categories of objects, like YSOs (region III); main sequence (region I/II); T Tauri and HAeBe stars (region III/IV), and planetary 
nebulae.
GL97  verified that about 2/3 of the planetary nebulae are found in the ``nebulae box" (region V), defined by Whitelock (1985).
The very likely post-AGB stars classified by Szczerba et al. (2007) are also plotted to illustrate the distribution
of evolved objects with single or double peak in their SED.

Bessell \& Brett (1988) studied the IR colours of Long-Period Variables (LPV), carbon stars and late-type supergiants. 
In Figure 8 we plot the schematic areas presented by Bessell \& Brett (1988, Fig. A3), where carbon-rich stars are enclosed 
by dashed lines and oxygen-rich LPVs fall in the area defined by continuous line. The post-AGBs of our sample are located in 
region III from GL97, beyond the right-hand-end schematic area from Bessell \& Brett. The only object coinciding with the region of 
carbon-rich stars is PDS551. 

\begin{figure}[!t]
\includegraphics[height=9cm,angle=270]{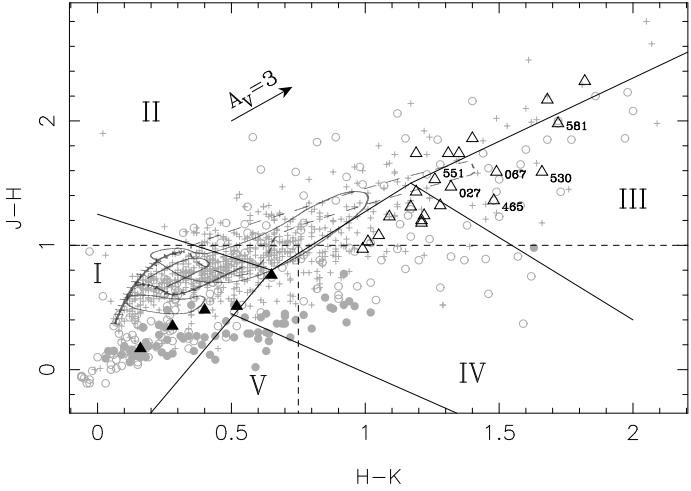}
\caption{ Near-IR color-color diagram  indicating the regions for different categories of objects defined by GL97 
(separated by thin full lines), Si\'odmiak et al. (2008) (thin dashed lines), and Bessell \& Brett (1988) (thick lines).
Triangles  are used to show our stars and grey circles represent {\it very likely} post-AGBs from the Toru\'n Catalog. 
Double peak SEDs are indicated by filled symbols (post-AGBs class IV). Grey crosses represent the HAeBe stars from {\it HBC}.}
\label{nir}
\end{figure}

It can  be noted in Figure 8 that several of our stars are found in  the same region as post-AGBs having single peak SED, 
while our objects having double peak SED appear in region I. 
As mentioned in Sect. 1.2, Si\'odmiak et al. (2008)  verified   in near-IR diagrams a clear separation of  nebulosity morphological classes
SOLE and DUPLEX .  The $J-H $ {\it vs.} $H-K$ diagram in Figure 8  shows
the regions occupied by DUPLEX ($J-H>$1) and SOLE  (lower left side, $H-K <$ 0.75). It is interesting to note that, in this diagram, the
{\it locus} of the ``stellar-like'' objects dubbed  by Si\'odmiak et al. (2008) (lower right side) coincides with the  region IV, expected for TTs and HAeBe 
stars (GL97).

\begin{figure}[!t]
\includegraphics[height=9cm,angle=270]{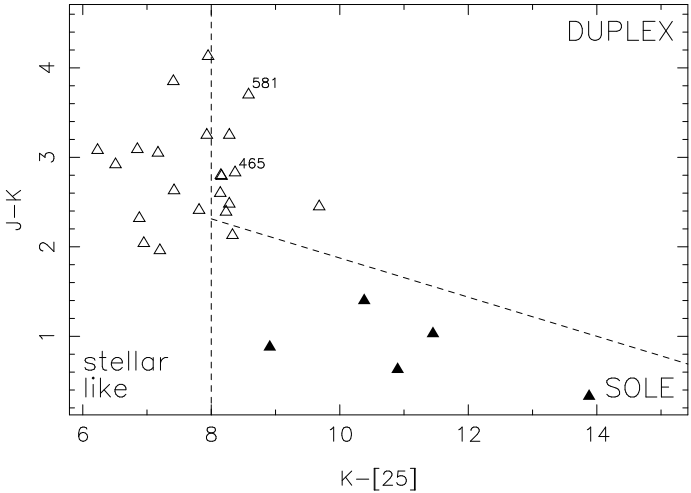}
\caption{Near- and mid-IR colors diagram 
indicating the regions defined by  Si\'odmiak et al. (2008) (dashed lines).
Filled triangles are used to show  those of  our stars with double peaked  SEDs, 
while open triangles represent single peak SEDs. }
\label{k25}
\end{figure}

Comparisons of SED shapes are also interesting to be discussed on basis of near- to mid-IR excess, which is related 
to the morphology of circumstellar structure (Ueta et al. 2000, 2007).
Figure 9 presents the $J-K$ {\it vs.} $K-[25]$ diagram where stellar-like post-AGBs are in the left part ( $K-[25]<$ 8), while
to the right are DUPLEXes (upper side) and SOLEs (lower part).
Si\'odmiak et al. (2008) also correlate the SED shape with morphological groups:
post-AGB class IV objects (double peak) are SOLE type, while classes I, II or III (single peak) are DUPLEX. 
As can be seen in Figure 9, the distribution of our double peak sources coincides with the SOLEs region. We
consider these sources, as well as those located in the DUPLEX region, possible being post-AGB stars.

In spite of the good correlations between near-IR and
SED shape, the discussion on elliptical or bipolar nebulosity types is only tentative since the morphology is unknown for
most of our objects. One exception is PDS204, for which
Perrin et al. (2009) used IR imagery and polarimetric data to
confirm the presence of an edge-on disk surrounded by an extensive 
envelope pierced by bipolar outflow cavities. The results of 
the present work indicate that PDS204 has characteristics similar to post-AGBs, in 
particular its circumstellar morphology inferred from J-K $\times$ K-[25] colors, which
is suggested to be DUPLEX (bipolar).


\section{Discussion and Summary of Results}

Aiming to analyse the similarities and differences between our sample of PDS sources and post-AGBs studied in other works, we have
established several criteria to diagnose if the selected objects are pre-MS or not. 
The main result of this comparative study is the nomination of the likely post-AGB stars, 
indicated by notes in Table 2. For each given note, the adopted criteria and respective values that we consider  typical 
characteristics of evolved objects are summarized as follows.

\noindent{\bf (a)} The possible association with  clouds was investigated by 
estimating the projected distance to the edge of the nearest cloud. According to the results presented in Sect. 3.1,  
note (a) indicates the objects having $d_{edge} > $ 0.5$^{\rm o}$.

\noindent{{\bf (b)}}
The circumstellar extinction, defined by $f_{\rm Av}$, was estimated by subtracting interstellar extinction, obtained from Dobashi et al.'s 
$A_V$ maps, of the total visual extinction, obtained from the $E(B-V)$  calculation. 
The fraction of source emission at 100$\mu$m, defined by 
$f_{\rm F100}$, was also evaluated in order to verify possible interstellar cirrus contamination. Note (b) indicates the objects 
showing  $f_{\rm Av} > $ 0.5 and $f_{\rm F100} > $ 0.5, as described in Sect. 3.3.

\noindent{\bf (c)}
Our sample was also analyzed according to the spectral features, in particular the equivalent width of H$_{\alpha}$ 
line. In Figure 4 the strength of  H$_{\alpha}$ is compared to the fractions of circumstellar emission $f_{\rm Sc}$ and $f_{\rm Av}$.
Note (c) indicates all the objects showing $W_{{\rm H}\alpha} >$ 50 {\AA} that have $f_{\rm Sc} >$ 0.87 and $f_{\rm Av} >$ 0.5.


\noindent{\bf (d)}
In Figure 5 (optical color-color diagram), seven objects have  $U-B >$ -1 and  $B-V <$ -0.2 that
we consider indicative of $U-B$ excess typical of post-AGBs. As discussed in Sect. 5.1, no difference on gravities could be verified 
in the case of high temperatures, but one object with $B-V >$ 0.1 seems to have low gravity, a possible indicator of evolved status.

\noindent{\bf (e)}
As mentioned in Sect. 5.2.1, the IRAS color-color diagram (Figure 6) shows eight of our objects with [12]-[25] $>$ 1,
all of them having F$_{12} > $ 2 Jy. This criterion has been suggested by van der Veen et al. (1989) to distinguish young stars, 
associated to clouds, from post-AGBs located in this part of the diagram that typically show F$_{12} > $ 2 Jy.  

\noindent{{\bf (f)} and {\bf (g)}}:
The distribution of our sample in diagrams of near-IR colors was 
checked according the suggestion 
by GL97 that establishes different regions of these diagrams for each evolutive post-AGB phase. The objects located 
in regions I and III of Figure 8 are indicated by note (f).
The diagram of near- and mid-IR colors shown in Figure 9 indicates that our objects having double peak 
SED are placed in the region of SOLE
type post-AGBs and single peak sources coincide with DUPLEX objects, as suggested by Si\'odmiak et al. (2008). 
According the results discussed in Sect. 5.2.3,  objects with K-[25]$>$ 8 have note (g).


\begin{table*}[ht] 
\caption{Parameters used in the sample analysis}
\smallskip
\begin{center}
{\scriptsize
\begin{tabular}{lcccccccc}
\hline
\noalign{\smallskip}
PDS &$f_{\rm Sc}$& cloud & d$_{edge}$& A$_V$& $f_{\rm Av}$ &$<F>$  &f$_{\rm F100}$& Notes \\
	&		&		&	$^o$	&	mag			&		&		Jy		&		&	\\
\noalign{\smallskip}
\hline
\noalign{\smallskip}
018	&	0.90	&	L1647	&	1.79	&	1.3	$\pm$	0.3	&	0.7	&	20	$\pm$	2	&	0.7	&	a,b	\\
027	&	0.96	&	L1659 	&	5.60	&	1.2	$\pm$	0.2	&	0.7	&	20	$\pm$	3	&	0.7	&	a,b,c,f,g	\\
037	&	0.93	&	FS134	&	-0.15	&	2.8	$\pm$	0.4	&	0.4	&	156	$\pm$	41	&	$<0.5$	&	e,g	\\
067	&	0.97	&	FS251	&	1.39	&	1.4	$\pm$	0.3	&	0.7	&	300	$\pm$	101	&	$<0.1$	&	a,f	\\
141	&	0.97	&	FS221	&	0.09	&	3.8	$\pm$	0.8	&	0	&	14	$\pm$	2	&	0.9	&		\\
168	&	0.75	&	L1536	&	-0.10	&	0.9	$\pm$	0.6	&	0	&	14	$\pm$	1	&	0.4	&		\\
174	&	0.97	&	L1616	&	0.07	&	1.2	$\pm$	0.2	&	0.6	&	6	$\pm$	4	&	1.0	&	b,c,d,e,g	\\
193	&	0.70	&	L1641	&	-1.02	&	2.6	$\pm$	0.6	&	0.4	&	37	$\pm$	16	&	0.7	&		\\
198	&	0.85	&	L1641	&	-0.76	&	2.6	$\pm$	0.6	&	0.3	&	37	$\pm$	16	&	$<0.6$	&		\\
204	&	0.95	&	L1570	&	5.91	&	0.9	$\pm$	0.4	&	0.7	&	17	$\pm$	1	&	0.9	&	a,b,c,e,g	\\
207	&	0.89	&	L1557	&	-0.28	&	0.3	$\pm$	0.5	&	0.9	&	16	$\pm$	2	&	0.8	&	b,g	\\
216	&	0.92	&	L1600	&	1.96	&	0.8	$\pm$	0.2	&	0.8	&	18	$\pm$	2	&	$<0.7$	&	a,b,c,d,g	\\
257	&	0.89	&	L1665	&	3.70	&	0.6	$\pm$	0.1	&	0.8	&	21	$\pm$	2	&	0.4	&	a	\\
290	&	0.94	&	FS114	&	1.27	&	1.6	$\pm$	0.2	&	0.2	&	37	$\pm$	9	&	0.5	&	a,f,g	\\
353	&	0.89	&	FS220	&	3.61	&	1.3	$\pm$	0.2	&	0.6	&	80	$\pm$	26	&	$<0.7$	&	a,b,c,d,g	\\
371	&	0.79	&	FS271	&	18.18	&	0.1	$\pm$	0.1	&	1.0	&	5	$\pm$	0	&	$<0.2$	&	a	\\
394	&	0.86	&	FS288	&	2.46	&	0.7	$\pm$	0.2	&	0.5	&	49	$\pm$	5	&	$<0.3$	&	a,e,g	\\
406	&	0.82	&	FS348	&	-0.20	&	1.2	$\pm$	0.5	&	0	&	35	$\pm$	5	&	0.2	&	d,g	\\
431	&	0.94	&	FS372	&	1.42	&	1.6	$\pm$	0.4	&	0	&	439	$\pm$	155	&	$<0.8$	&	a,e,f,g	\\
465	&	0.92	&	L273	&	0.58	&	0.6	$\pm$	0.2	&	0.8	&	30	$\pm$	3	&	0.7	&	a,b,c,d,e,f,g	\\
477	&	0.87	&	L318	&	0.10	&	2.6	$\pm$	0.4	&	0.4	&	120	$\pm$	14	&	0.2	&	d	\\
518	&	0.91	&	L511	&	0.09	&	5.4	$\pm$	1.0	&	0.2	&	105	$\pm$	11	&	0.9	&	d	\\
520	&	0.84	&	L571	&	0.49	&	5.5	$\pm$	0.8	&	0	&	79	$\pm$	26	&	$<0.1$	&		\\
530	&	0.82	&	L637	&	0	&	1.0	$\pm$	0.3	&	0.3	&	38	$\pm$	5	&	0.3	&	f	\\
543	&	0.94	&	L621	&	-3.01	&	4.6	$\pm$	0.6	&	0.2	&	164	$\pm$	47	&	0.2	&	e,f,g	\\
551	&	0.96	&	L628	&	0.30	&	6.3	$\pm$	0.7	&	0	&	196	$\pm$	43	&	$<0.5$	&	f,g	\\
581	&	0.94	&	L813	&	1.28	&	0	$\pm$	0.1	&	1.0	&	21	$\pm$	3	&	0.8	&	a,b,c,d,e,f,g	\\
\hline
\end{tabular}
}
\end{center}
Columns description: (1)  PDS name; (2) fraction of circumstellar luminosity; (3, 4) nearest cloud and respective distance to the edge;
 (5) visual extinction from Dobashi et al. (2005); (6) ``net" visual extinction;  (7,8) background mean flux and fraction of source emission at 100$\mu$m.\\
Notes: (a) $d_{edge}>$0.5$^{\rm o}$; (b) $f_{\rm  Av} >$ 0.5 and $f_{\rm F100} >$ 0.5; (c) high values of $W_{{\rm H}\alpha}$,
$f_{\rm  Av} $ and $f_{\rm F100}$; (d)  U-B excess; (e) [12]-[25] $>$1 and $F_{12} >$ 2 Jy;
near-IR colors typical of post-AGB: (f) J-H and H-K, (g) J-K and K-[25].
\end{table*}

\section{Conclusions and Perspectives for future works}


The whole set of adopted criteria suggests that 26\% (7/27 sources) of the studied sample  are very likely post-AGB stars, 
which achieved five or more notes. Four objects having three or four notes are considered possible post-AGBs. 
As discussed below, PDS371 is also included in this list indicating that 18\% (5/27) of the sample possibly are post-AGBs. 
Among the other objects having less than three notes, 8/27 sources (30\% of the sample) are unlikely post-AGB since 
they were confirmed in the literature as pre-MS stars. The nature of the remaining 7/27 objects is unclear.
Individual comments are given in Appendix A to present recent results from literature
related to  the nature of our stars.

PDS465 and 581 are well known post-AGBs. The other very likely post-AGBs indicated by us are PDS~27, 174, 204,
216, and 353. Results from other works support these suggestions at least for two of them: PDS~27 (Su\'arez et al. 2006) and 
PDS 174 (Sunada et al. 2007).

The unlikely post-AGB objects: PDS~18, 37, 141, 168, 193, 198, 406, and 518 are confirmed pre-MS stars (see Appendix A). 
Other sources with less than three notes are: PDS~67, 207, 257, 477, 520, 530, and 551, which were not previously studied in the literature.

Other objects whose nature remains to be confirmed are PDS~290, 394, 431, 543 that we suggest to be possible post-AGB objects.
As discussed in Section 4, PDS371 is included in this group since it shows spectral features similar to proto-PN.
The {\it AKARI} colours, discussed in Sect. 5.2.2,
were not used to nominate the post-AGBs, due to the position of our sample, mainly concentrated in the region of overlap of
young and evolved objects in Figure 7. However, it is interesting to note that three sources appear out of this overlap region.
Indeed, PDS 204 (very likely post-AGB), and PDS 394 and 543 (possible post-AGBs) have colours [9]-[18] $>$ 3 mag,
which coincide with the {\it locus} of most of the post-AGB class IV, confirming the classification
that we suggest for these PDS sources.

We are aware that additional observations 
are needed to definitively classify the candidates, as mentioned by Szczerba et al. (2007) in their evolutive
catalogue of post-AGB  objects.  
Several techniques and instruments operating in different spectral domains can be used to confirm the evolved 
nature of our candidates. Detailed spectroscopic analysis is required to obtain a better determination of spectral 
type and luminosity class that could be evaluated through the spectral lines sensitive  to changes in gravity (Pereira
\& Miranda 2007). Spectral synthesis, for example, would provide abundances of C and O, whose ratio is used in 
the classification of post-AGB  stars (van der Veen et al. 1989). Abundances of Ba and Sr are also interesting to be 
determined, since the enrichment  of elements produced by s-process is expected to occur during the post-AGB 
phase (Parthasarathy 2000). However, 
a detailed spectral analysis of absorption lines is constrained to objects with intermediary to low $T_{eff}$, which is 
not the case of our sources that do not show these spectral features. It is more indicated, in this case, to study the emission line 
profile of  He I $\lambda$ 5876,  or  the strength of [N II] $\lambda$5754 relative to [O I] $\lambda$ 6302, 
that  were used by Pereira et al. (2008), for example, to indicate the nature of a proto-PN candidate. 
Evidence of a carbon-rich chemistry can be also 
obtained through mid-infrared spectroscopy  from 10 to 36 $\mu$m, as the example of the study by Hrivnak et al. (2009).

Direct imagery remains the most interesting tool to reveal the morphology of the nebulae. Our sources show several 
indications of having a considerable amount of circumstellar material that could be detected in high resolution images. 
Ueta et al. (2007), for example, studied the post-AGB circumstellar structures via dust-scattered linearly polarized starlight, 
by using  imaging-polarimetric data obtained with NICMOS/HST. It would be very interesting to develop a SED fitting 
based on more reliable circumstellar parameters and a disk model more realistic than that adopted in the present work. 
We intend to apply a radiative transfer model specifically developed to study the physical conditions like mass loss and 
envelope density of the post-AGB.

Our results indicate new likely and possible post-AGBs  that open promising and exciting 
perspectives to continue  studying these targets.

\begin{acknowledgements}
RGV and JGH thank support from FAPESP (Proc. No. 2008/01533-4; Proc. No. 2005/00397-1). 
This research has made use of the SIMBAD database, operated at CDS, Strasbourg, France.
\end{acknowledgements}

\begin{appendix} 

\section{Individual Comments}

The literature has been searched for results confirming the nature of the 
studied objects, which are found for 15/27 of the sample, all of them 
consistent with the results of the present work. In this section we separate 
these references in three parts: (i) previously known young stars; 
(ii) very likely post-AGBs; (iii) possible post-AGB or pre-MS.

\subsection{Young stars}

{\bf PDS18} is included in the study of YSOs showing IR nebula, presented by 
Conneley et al. (2007).  A 
reflection nebulae associated to PDS18 is catalogued by Magakian (2003). 
Additionally, Conneley et al. (2008) also verified the presence of an 
arc-shaped nebula ($\sim$ 8 arcsec to the northeast) around this object.

{\bf PDS141} (DK Cha), one of the most studied object of our sample, is a Class I 
pre-MS star. Results based on {\it Spitzer} data have been presented by Porras et al. 
(2007), Alcal\'a et al. (2008), Ybarra \& Lada (2009), Evans et al. (2009), for 
example. The disk structure was studied by Liu et al. (2007). 
CO emission has been reported by Otrupcek et al. (2000).

{\it Spitzer} data are also provided for {\bf PDS168} (Luhman et al. 2006, Furlan et al. 
2008), which is considered a Class I protostar (White \& Hillenbrand 2004, 
Terebey et al. 2009). The protoplanetary disk was studied from H$_2$ 
emission (Bitner et al. 2008) and from millimeter data (Andrews \& Williams 
2005).

{\bf PDS193} is in the 3$\mu$m spectroscopic survey of HAeBes, presented by Acke 
\& van den Ancker (2006). {\bf PDS406} was included in the study of Class 0 
protostars by Froebrich (2005) and appears in the reflection nebulae 
catalogue from Magakian (2003). 

Sunada et al. (2007) included {\bf PDS198} in their 
study of H$_2$O maser, but no maser emission was found for this object, 
probably due to its intermediate-mass (F0 type star). Acke \& van den 
Ancker (2006) consider PDS198 as HAeBe.

{\bf PDS518} is also a well studied young B star (Acke et al. 2008), for which 
3$\mu$m spectroscopy of protoplanetary disks is presented by Goto et al. 
(2009); the circumstellar structure was also studied by Alonso-Albi et al. 
(2009). Conneley et al. (2008) included PDS518 in a study of multiplicity of 
embedded protostars.

\subsection{Post-AGB}

{\bf PDS27} appears in the list of Su\'arez et al. (2006) who presented a 
Spectroscopic Atlas of post-AGB stars. CO emission has been reported by Urquhart et al. (2007).

The emission of H$_2$O maser (usually found in massive protostars) was 
not detected in the survey by Sunada et al. (2007) in the case of {\bf PDS174} that 
is a result consistent with our suggestion of post-AGB candidate for this object. 

As stated before, PDS465 and PDS581 are confirmed post-AGB objects largely 
studied in the literature. The most recent results for {\bf PDS465} are presented by 
Lee et al. (2009), De Marco (2009), Luna et al. (2008), Raga et al. (2008), 
among others, and for {\bf PDS 581}: Steffen et al. (2009), Montez et al. (2009), 
Dennis et al. (2008), Sanchez Contreras et al. (2008), for example.
CO emission associated to PDS581 has been reported by Urquhart et al. (2008).
Both objects are listed in the Toru\'n Catalogue, classified by Szczerba et al (2007)
as {\it IRAS} selected source (PDS465) or reflection nebulosity (PDS581).

{\bf PDS 394} is  studied as post-AGB 
candidate by Su\'arez et al. (2006) and Szczerba et al. (2007). 

{\bf PDS204} (MWC778) is considered a peculiar object in the 
H$_2$O maser study of post-AGB (Su\'arez et al. 2007, 2006).
CO emission has been reported by Kawamura et al. (1998) and Urquhart et al. (2007).
Herbig \& Vacca (2008) suggested that MWC778 has a SED slope of pre-MS 
embedded object, which has spectral type F or G, due to the presence of metallic 
absorption lines. 
From IR imagery and polarimetric data, Perrin et al. (2009) 
confirm the presence of an edge-on disk surrounded by an extensive 
envelope pierced by bipolar outflow cavities. They argue that MWC778 has 
high bolometric luminosity inconsistent with F or G type star. The results of 
the present work indicate that PDS204 has characteristics similar to post-AGBs, in 
particular its circumstellar morphology inferred from J-K $\times$ K-[25] colors, 
which are found in DUPLEX (bipolar) post-AGBs. 

{\bf PDS216} has been included in the study of hydrocarbon molecules in Herbig Ae stars 
(Acke et al. 2010) aiming to verify the correlation of effective temperature with
the 7.8 $\mu$m feature, when comparing their sample with low- and high-mass stars.
Considering that similar results are also found for S-type AGB stars (Smolders et al. 2010),
Acke et al. (2010) conclude that the chemistry of hydrocarbon molecules in the circumstellar 
environment is mainly affected by stellar radiation field  regardless of the evolutionary status.

\subsection{Possible post-AGB or pre-MS}

{\bf PDS37} is in the list of post-AGB stars studied by Szczerba et al. (2007), but 
it is considered YSO in the work of Mottran et al. (2007) and in the {\it Spitzer}
mid-IR survey by Carpenter et al. (2008).
CO emission has been reported by Urquhart et al. (2007).

{\bf PDS371} has been included in the photometry and spectroscopy for luminous
stars catalogue by Reed (2005). However, no additional information is available than those
given by the PDS Catalogue (Vieira et al. 2003).

CO emission has been reported by Brand et al. (1987) for PDS290 and PDS353.

Pestalozzi et al. (2005) detected in {\bf PDS431} methanol maser emission, which 
usually is considered an indicator of massive star formation as well as 
observed in Class 0 protostars. CO emission has been reported by Urquhart et al. (2007).

\end{appendix}
\end{document}